# Performance Estimation of 2*4 MIMO-MC-CDMA Using Convolution Code in Different Modulation Technique using ZF Detection Scheme

Mr. Atul Singh Kushwah, Ms. Priya Rathore & Mr. Ramsewak Kumar

*Asst. Professors & Electronics & Communication & Indore Institute of Science & Technology-II, Indore (M.P), India*

*Abstract -* **In this paper we estimate the performance of 2*4 MIMO-MC-CDMA system using convolution code in MATLAB which highly reduces BER. MC-CDMA (Multi Carrier Code Division for Multiple Access) is a multi-user and multiple access system which is formed by the combination of OFDM and CDMA and convolution encoding scheme is used in encoder of CDMA as FEC (Forward Error Correction) code to reduce BER (Bit Error Rate). MC-CDMA system is a multi-carrier system in which single wideband frequency selective carrier is converted into parallel narrowband flat fading multiple sub-carriers to optimize the performance of system. Now this system further improved by combination of 2*4 MIMO (Multiple Input Multiple Output) system which utilizes ZF (Zero Forcing) decoder at the receiver to reduce BER and also ½ rate convolutionally encoded Alamouti STBC (Space Time Block Code) block code as transmit diversity of MIMO for multiple transmission of data through multiple transmit antenna. Main advantage of using MIMO-MC-CDMA using convolution code is to reduce the complexity of system and to reduce BER with increased gain. In this paper we analyze system performance in different modulation schemes like, QPSK, 8-PSK, 8-QAM, 16-QAM, 32-QAM and 64-QAM in Rayleigh fading channel using MATLAB.**

## I. INTRODUCTION

Due to increased demand of high data rate and low probability of error in this paper we utilizes the technique of MIMO, CDMA and OFDM results enhanced technique for minimizing error rate. MC-CDMA is multicarrier and multiple access system which is a combination of OFDM and CDMA [10]. This MC-CDMA forms multiple accesses and multiple carrier system in frequency selective flat fading channel. The MC-CDMA results improved efficiency of the wireless communication system which results high data rate and low probability of error.

In this paper MIMO is combined with MC-CDMA to increase throughput. MIMO is multiple antenna system in which multiple receive diversity and multiple transmit diversity i.e half-rate convolutionally encoded Alamouti STBC code is used for synchronization of system to reduce ISI. To detect signal orthogonality ZF detection scheme is used. And as a result MIMO-MC-CDMA [1] is formed by all above operations using MATLAB is then it will be encoded using convolution code as FEC encoder. This MIMO-MC-CDMA using convolution code then analyzed in *QPSK, 8-PSK, 8-QAM, 16-QAM, 32-QAM and 64-QAM* modulation techniques in Rayleigh fading channels.

## II. MULTI CARRIER CODE DIVISION MULTIPLE ACCESS (MC-CDMA)

MC-CDMA [2,3,4] is a combination of system of OFDM and CDMA system [11]. This system allows the multiple users to access the wireless channel simultaneously by modulating and spreading their input data signals in frequency domain using PN spreading sequences. MC-CDMA combines the multipath fading of OFDM system with the multi-user access of CDMA system.

### A. Mathematical model

In present MC-CDMA system we assume that signal detection scheme will be implemented for two transmitting and four was receiving antennas. Each user use single transmitting antenna system. Suppose H represent a channel matrix as $h_{ij}$ for the channel gain between the i-th transmitting antenna and the j-th receiving antenna, j=1,2,3,4 and i=1,2. Each user data and the equivalent received signals are represented by $a=[a_1, a_2]^T$ and $y=[y_1\ y_2\ y_3\ y_4]^T$ respectively, in which $a_i$ and $y_j$ denote the transmit signal from i-th transmitting antenna and the received signal at the j-th received antenna respectively. Assume $n_j$ denote the white Gaussian noise of variance of $\sigma_{n4}$ at the j-th receiving antenna and $h_i$ denote the i-th column vector of the channel matrix H. Now received signal y for system can be represented by

$y=Ha+n$ …………….……(1)

Where, $n= [n_1,n_2,n_3,n_4]^T$

As the interference signals from other transmitting antennas are reduced for detecting the desired signal from the target transmitting antenna, the detected desired signal from the transmitting antenna by inverting channel effect from a weight matrix W is

$\tilde{A}=[\tilde{a}_1\ \tilde{a}_2]=W_y$……….(2)

For Zero-Forcing (ZF) scheme, the ZF weight matrix is represented by

$W_{ZF}=(H^H H)^{-1} H^H$ …….(3)

where $()^H$ denotes the Hermitian transpose and the detected desired signal from the transmitting antenna using the following relation

$\tilde{a}_{ZF}=W_{ZF} y$ ………(4)

It is intended to find the transmitted signal vector by Sphere Decoding (SD) scheme with minimum least ML metric. Suppose $y_R$ and $y_I$ represents the real and imaginary parts of the received signal y, i.e. $y_R$= Re{y} and $y_I$= Im{y}. Corresponding to that the input signal $x_i$ and the channel gain $h_{ij}$ from ith transmitting antenna to jth receiving antenna can be showed by $a_{iR}$ = Re{$a_i$} and $a_{iI}$ = Im{$a_i$} and $h_{ijR}$ = Re{$h_{ij}$} and $h_{ijI}$ = Im{$h_{ij}$} respectively. We can say that,

$$\hat{y}=[y_R \ jy_I]^T \quad \ldots\ldots(5)$$

From (6), the detected desired signal $\tilde{a}_{SP}$ with its real and imaginary components from the transmitting antenna can be represented by [3]:

$$\tilde{a}_{SP}=[\tilde{a}_{1R} \ \tilde{a}_{2R} \ \tilde{a}_{1I} \ \tilde{a}_{2I}]^T =[\ \hat{H}^T \ \hat{H} \ ]^{-1}*[\ \hat{H}^T \ \hat{y} \ ]\ldots(6)$$

### III. MULTIPLE INPUT MULTIPLE OUTPUT (MIMO)

MIMO systems [5] use multiple transmit antennas and multiple receive antennas at the receiver, so both transmit and receive diversity schemes are applied to reduce fading resulting from signal variations by wireless channel. It depends on the degree at which the multiple data replicas are faded independently; the system provides diversity gains which represents the difference in SNR at the output of the diversity combiner as compared to that of single branch diversity at certain probability level.

A MIMO system consisting of *N* transmit antenna elements equal to two, and of *M* receive antenna elements equal to four was modeled, accordingly diversity order of 4 can be achieved. For improving the performance combination of the multiple versions of the signals created by different diversity schemes is required. This paper applies Zero Forcing (ZF) decoder to combine M received signals to resonate on the mostly required desired transmitted signals. The sum of the received SNRs form M different paths is the effectively received SNR of the system with Alamouti STBC of 2*4 diversity. The receiver required to demodulate all M receive signals in case of ZF for a source with independent signals in the received antennas.

### IV. CONVOLUTION CODE

Non-systematic convolution encoder is shown in fig.1. The data were not present at the output of the encoder and were replaced by modulo 2 sum of the data at instant i is $d_i$ and the data at instants i − 2 is $d_{i-2}$ and the data at instant i − 3 is $d_{i-3}$ and so on. The rate of the encoder remains is ½ i.e. 2 encoded output bits for each input bits.

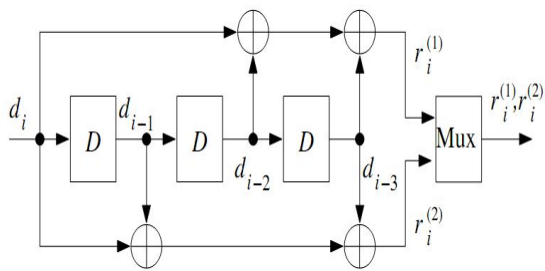

Fig. 1. Non-systematic convolutional encoder [12].

### A. Generic representation

Shift registers, XOR gates and v flip-flops are used for the construction of non-systematic convolution encoder. Now we define the main characteristic of convolutional codes i.e. the constraint length, here equal to v +1. The register at instant i is characterized by the v no. of bits $s_i^{(1)}, s_i^{(2)},...,s_i^{(v)}$ then states of v bits are also represented in the form of a vector $s_i = (s_i^{(1)}, s_i^{(2)}, \cdots, s_i^{(v)})$. There are $2^v$ possible state values in convolution encoder so that we can often denote in natural binary or binary decimal form. So the state of encoder is $2^3$ which is determined by no. of flip flops which is determined by $2^3$=8. Assume $s_1$ =1, $s_2$ =1 and $s_3$=0, then the encoder is in state 110 in natural binary, i.e. 6 in decimal form.

Using the m no. of coefficients $a_j^{(l)}$, used to form a vector $d_i$ is selected then addition is done with the previous content of flip-flops but not in case of the first flip-flop, which forms the value to be stored in the following flip-flops. New content of a flip-flop depends on the current input and previous content of flip-flop. In case of $b_j$ null coefficients, the resultant input depends only on the sum of the the components selected of $d_i$. In case when $b_j$ coefficients are non-zero then these components are added with $d_i$ and the recursive code is generated. So the successive states of the registers depends on previous inputs and present input through the flip-flop which finally produces redundant components $r_i$ which is produced by the summation of the content of flip-flops by coefficients g.

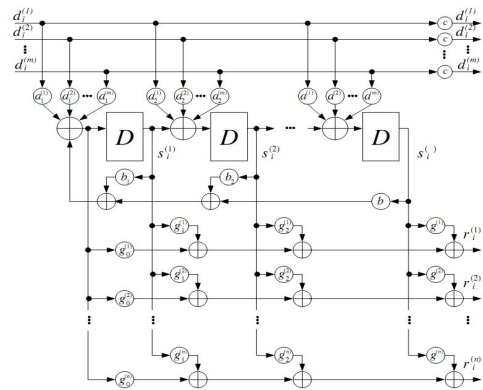

Fig.2. Generic representation of an encoder for convolutional codes [12].

In this paper we are using non-systematic ½ rate convolution encoder as FEC at the MIMO-MC-CDMA in which 1040 message bits are encoded into 2080 encoded bits i.e two output bits are formed in each message input bits.

### V. MIMO-MC-CDMA COMMUNICATION SYSTEM MODEL USING CONVOLUTION CODE

Communication system model of MIMO-MC-CDMA using convolution is shown in fig.3.

In this communication system we are assuming transmitter sending random sequence to the receiver so we are using random PN sequence generator using MATLAB. Now convolution encoding is done as FEC technique to reduce error probability and spreading of sequence is done using PN sequence generator. Then different modulation scheme is used like *QPSK, 8-PSK, 8-QAM, 16-QAM, 32-QAM and 64-QAM* this is represented by modulator block. Previously described system is MC-CDMA system which is already described in section 2 as Multi-Carrier Code Division Multiple Access (MC-CDMA). Now MIMO encoder half-rate convolutionally encoded Alamouti's STBC block code is used which will be described in section 3 as Multiple Input Multiple Output (MIMO). Combination of MIMO and MC-CDMA forms MIMO-MC-CDMA using convolution code as shown in fig.1. Now signal is then transmitted through Rayleigh Fading Channel [6]. Then receiver receive the signal in reverse fashion with ZF decoder for the recovery of transmitted signal and BER calculation is done for estimating the system performance. In MIMO system two transmit antenna and four receive antenna is used. In this paper we are sending message bits which is random in nature or depend on user then this data is moved through FEC encoder, here ½ rate convolution encoder having constraint length of 3 is used so after this block no. of bits will be double, then this encoded bits are spreaded using PN sequence which forms 8 bits in each input bits i.e. input bits*8 then resultant bits are formed after the spreading of encoded sequence. Then these encoded and spreaded sequences of bits are passed through modulator which is modulated depends on the type of modulation used. Then these modulated data is reshaped into parallel form for OFDM then IFFT is done which convert frequency selective wide-band carriers into parallel flat-fading narrowband carriers which are orthogonal in nature then this data reshaped to parallel to serial then CP addition is done to remove the effect of ISI which completes the process of OFDM then this data then transmitted through MIMO encoder which use Alamouti STBC block code for 2 transmit diversity and 4 receive diversity techniques in which 4*2 channel matrix is formed by using this MIMO diversity, then ZF detection scheme is used at the receiver then reverse process is done for receiving the input bits.

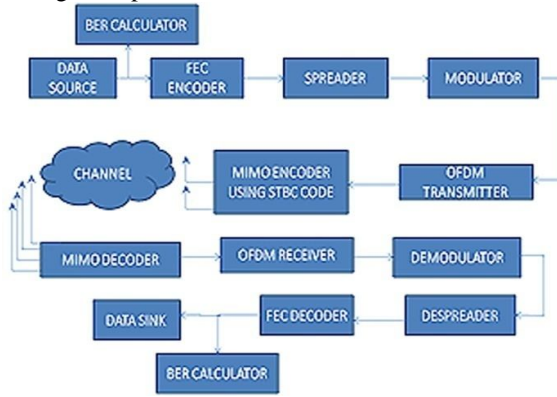

Fig.3. Communication System Model OF 2*1 MIMO-MC-CDMA using convolution code

VI. SIMULATION RESULTS AND DISCUSSION:

Table 1 shows the simulated model parameters of MIMO-MC-CDMA [6,7,8,9] using convolution code in mentioned different modulation technique.

Performance estimation of MIMO-MC-CDMA using convolution code is shown in fig.4-9.

Fig. 9 shows the comparative analysis of different modulation schemes in MIMO-MC-CDMA using convolution code.

Table 2 shows the performance estimation of all modulation schemes in terms of gain and BER.

TABLE:1. SIMULATED MODEL PARAMETERS.

| FEC Encoder | Convolution encoder |
|---|---|
| Channel Encoder | ½ rate convolution encoder |
| Signal detection scheme | Zero forcing |
| Channel | Rayleigh Fading Channel |
| Signal to Noise Ratio | -10dB to 20 dB |
| CP Length | 1280 |
| OFDM Sub-carriers | 6400 |
| No. of transmitting and receiving antennas | 2*4 |
| Modulation Schemes | QPSK, 8-PSK, 8-QAM, 16-QAM, 32-QAM and 64 QAM |

From table.2 and Fig.10 we can say that QPSK shows high gain (17.56 dB) and very low BER with respect to other modulation schemes in -5dB SNR. This is possible by using convolution code as FEC encoding scheme in MIMO-MC-CDMA errors in QPSK and 8-QAM is zero which shows very low probability of error in system.

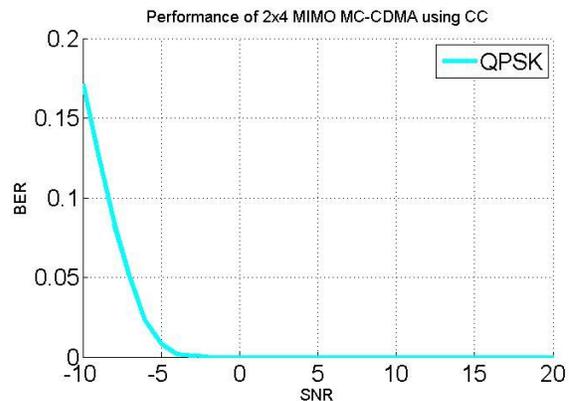

Fig.4. Performance estimation of MIMO-MC-CDMA using convolution code in QPSK modulation scheme.

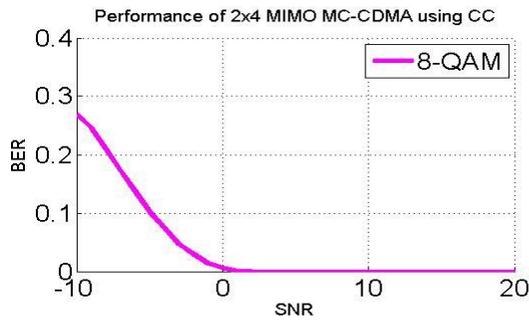

Fig.5. Performance estimation of MIMO-MC-CDMA using convolution code in 8-QAM modulation scheme.

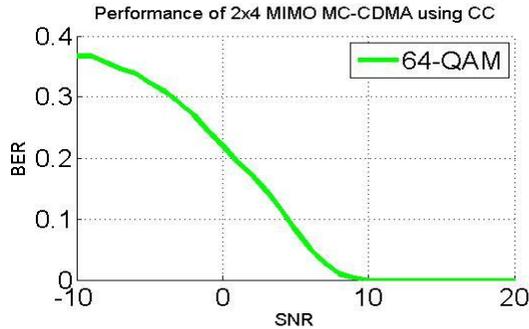

Fig.6. Performance estimation of MIMO-MC-CDMA using convolution code in 64-QAM modulation scheme.

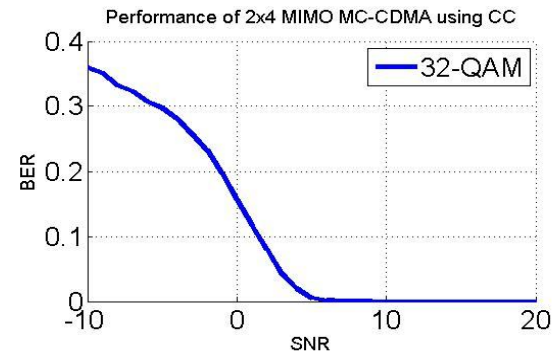

Fig.7. Performance estimation of MIMO-MC-CDMA using convolution code in 32-QAM modulation

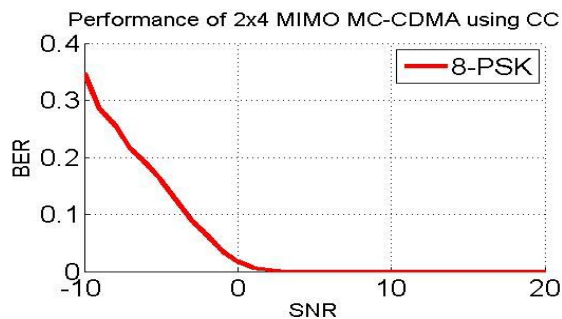

Fig.8. Performance estimation of MIMO-MC-CDMA using convolution code in 8-PSK modulation

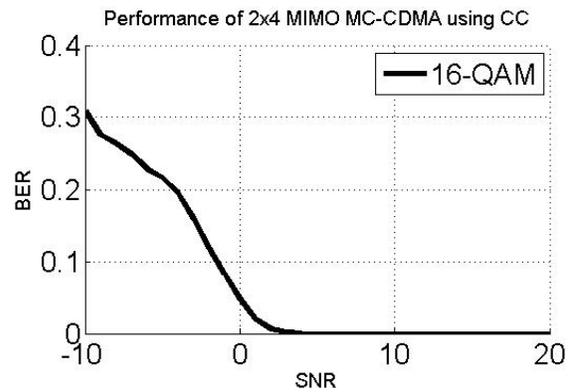

Fig.9. Performance estimation of MIMO-MC-CDMA using convolution code in 16-QAM modulation

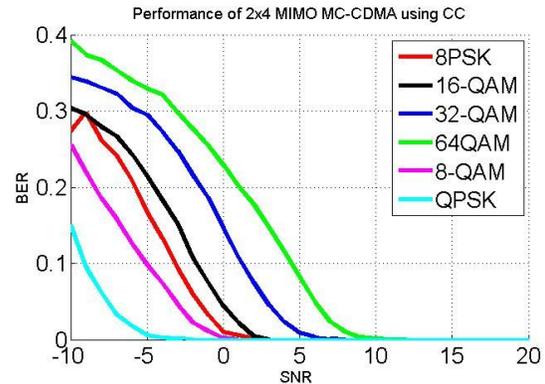

Fig.10. Performance estimation of MIMO-MC-CDMA using convolution code in 8-QAM, 16-QAM, 32-QAM, 64-QAM, 8-PSK and QPSK modulation scheme.

**TABLE: 2. PERFORMANCE ESTIMATION AT -5DB SNR WITH RESPECT TO 64-QAM MODULATION TECHNIQUE AS SHOWN IN FIG.10:**

| *Modulation* | *BER at -1dB* | *Gain w.r.t 64-QAM* |
|---|---|---|
| QPSK | 0.005769 | 17.56dB |
| 8-QAM | 0.09821 | 5.25dB |
| 8-PSK | 0.1672 | 2.946dB |
| 16-QAM | 0.2144 | 1.867dB |
| 32-QAM | 0.2946 | 0.486dB |
| 64-QAM | 0.3295 | 0dB |

VII. CONCLUSION

Fig.10 shows the comparative analysis of MIMO-MC-CDMA using convolution code in QPSK, 8-PSK, 8-QAM, 16-QAM, 32-QAM and 64 QAM modulation schemes. Table 2 shows the comparative estimation of different modulation schemes which shows that as modulation scheme order is higher results increase in BER. This paper aims to reduce bit error rate which is found in QPSK modulation scheme with gain of 17.56 dB with respect to 64-QAM which represents that the gain of QPSK is higher as compared to other modulation technique with very low probability of error because errors were removed at 0dB in QPSK. For 3G and 4G communication system 64-QAM modulation technique is preferred which contain BER up to 9dB, i.e. errors are removed in 64-QAM at 9dB SNR which is highly optimized by using MIMO-MC-CDMA using convolution code.